\documentstyle[aps,epsfig,float,twocolumn,prl]{revtex} 
\newcommand{\be}{\begin{equation}}
\newcommand{\ee}{\end{equation}}

\begin{document}
\twocolumn[\hsize\textwidth\columnwidth\hsize\csname @twocolumnfalse\endcsname 
\draft
\title{Exact Multifractal Spectra for Arbitrary Laplacian Random Walks}
\author{M. B. Hastings}
\address{
Center for Nonlinear Studies and Theoretical Division, Los Alamos National
Laboratory, Los Alamos, NM 87545, hastings@cnls.lanl.gov 
}
\date{August 31, 2001}
\maketitle
\begin{abstract}
Iterated conformal mappings are used to obtain exact multifractal spectra of 
the harmonic measure for arbitrary Laplacian random walks in two dimensions.
Separate spectra are found to describe scaling of the growth measure in time, 
of the measure near the growth tip, and of the measure away from the growth
tip.  The spectra away from the tip coincide with those of conformally
invariant equilibrium systems with arbitrary central charge $c\leq 1$, with $c$
related to the particular walk chosen, while the scaling in time and near
the tip cannot be obtained from the equilibrium properties.
\vskip2mm
\end{abstract}
]
Diffusion-limited aggregation\cite{dla} and the dielectric breakdown 
model\cite{dbm}(DBM) are central problems in fractal growth which still
pose difficult problems.  As the growth is controlled by a Laplacian 
field, the harmonic measure on these clusters is closely connected to 
the dynamics.  Quantitative characterization of the measure is provided
by the multifractal exponents\cite{mf}.

There are unfortunately only a few exact results for these exponents.
The Makarov scaling law\cite{makarov} is true for any simply connected
curve, and the electrostatic scaling law\cite{electro} is exact for
the DBM.  In an important recent advance, exact exponents were found for 
systems that may be described using conformal field 
theory\cite{duplantier}(CFT); a series of spectra were found, labelled
by the central charge $c$ of the theory.  This includes both equilibrium
statistical mechanics clusters, such as Ising clusters at $c=1/2$, as
well as certain growth processes, such as random walks at $c=0$.

However, these exponents for growth processes only describe the static
properties of these systems.  Interesting dynamical questions,
such as fluctuations about the average growth rate, remain unanswered 
in the absence of exact results.  To address this, we will consider a 
nonequilibrium growth process, the Laplacian random walk\cite{lrw}(LRW), and 
show that its static properties away from the growth tip coincide with
those of CFT clusters,
but we will also obtain nontrivial scaling near the growth tip, giving
rise to nontrivial dynamical scaling.  
This should shed light on more complicated growth
processes such as DLA where the statics and dynamics are so deeply
interwoven.

The LRW is defined by a growing walk on a lattice,
in which at every stage of growth a Laplacian field is computed with boundary
conditions such that it vanishes on the walk and grows logarithmically at
infinity.  From this field an electric field $E$ is computed, and the
probability of selecting any site near the tip for growth is proportional
to $E^{\eta}$ at that site, $\eta>0$.  This can
be considered as a dielectric breakdown model in which growth occurs only
at the tip.

In this paper, we propose a model of iterated conformal maps for the LRW.
This model is not conformally invariant at the microscopic level,
having a fixed particle size; we then show that 
at large scales this model is equivalent to another discrete
model which possesses microscopic conformal invariance.  
We obtain the exact multifractal spectrum
as a continuously varying function of a parameter $m$, which will be indirectly
connected to the $\eta$ parameter above.  The $\eta=1$ LRW
corresponds to $m=1/2$ or a $c=-2$ CFT\cite{lrexact}.  For other LRWs, 
there is a continuous relation between $m$ and $c$ for exponents of the
measure away from the growth tip.  While our
model has the same continuum limit as the stochastic Loewner
equation\cite{sle}, we use the discreteness to obtain
additional exponents for scaling in time and measure near the growth tip.

{\it Conformal Mappings for the LRW ---}
We use the method of iterated conformal mappings\cite{hl} to construct
an off-lattice LRW argued to be in the same universality class
as that defined above.   In the LRW,
average growth follows the direction of the electric field at the
tip of the walk, but with fluctuations about this preferred
direction.  We will construct a model with the same features.

Let $F(z)$ be a map from the real line to a growing walk, with $F$ analytic
in the upper half plane.  Define $f_{x,\lambda}(z)$ to produce a bump of
linear size $\sqrt{\lambda}$ at point $x$ on the real line.  For
example, $f_{0,\lambda}=\sqrt{z^2-\lambda}$.  If $F(x)$ is the
tip of the walk, then $F(f_{x,\lambda}(z))$ maps to the walk grown
by a distance $|F'(z)|\sqrt{\lambda}$ in the direction of the
field at the tip.  Thus we propose the following model: fix the growth
tip to be the image of $z=0$.
At each growth step, first compose 
$F(z) \rightarrow F(f_{0,\lambda}(z))$, to grow the walk, where
$\lambda=|F'(0)|^{-2} l_1^2$ with
$l_1$ a length that
determines the size of the growth step in physical space.  Then 
compose $F(z) \rightarrow F(z\pm \sqrt{\lambda} l_2/l_1)$, 
where the plus or minus
sign is chosen randomly.  This will shift the growth tip by a distance
$l_2$ away from the preferred growth direction; the greater the ratio
$l_2/l_1$ the higher will be the dimension of the random walk and so the
smaller the effective $\eta$.   

We wish to determine the average of the $q$-th moment of $|F'(z)|$
at a point $z=x+iy$; that is, $\langle |F'(z)|^{q} \rangle$, where the
brackets denote ensemble averaging.    This describes the field
as a function of cutoff in the {\it mathematical plane, $z$}, which we
we will later convert to scaling in the {\it physical plane}.  
Denote this average after $n$ growth steps
as $\rho_q^n(x,y)$.  By the chain rule we have
\begin{eqnarray}
|F(f(z))'|^q=
(1+\frac{\lambda}{2} q \frac{x^2-y^2}{|z|^4})
|F'(z-\frac{\lambda}{2z}\pm \sqrt{\lambda} l_2/l_1)|^q,
\end{eqnarray}
where we have used the approximation $f(z)=z-\lambda/(2z)$, valid at scales
larger than $l_1$.

Taking a continuum limit in the number of growth steps
gives the differential 
equation
\begin{eqnarray}
\label{de}
\frac{\partial \rho_q(x,y)}{\partial n}=\frac{\lambda}{2}
\Bigl(m \partial_x^2 -\frac{x}{x^2+y^2}\partial_x+\nonumber 
\frac{y}{x^2+y^2}&\partial_y \\+
q\frac{x^2-y^2}{(x^2+y^2)^2}
\Bigr) \rho_q(x,y),&
\end{eqnarray}
where we have replaced the discrete translation $x\rightarrow
x\pm \sqrt{\lambda}l_2/l_1$ by a diffusion term with coefficient
$m=(l_2/l_1)^2$, which defines the ``meander" of the walk.

Recall that $\lambda=|F'(0)|^{-2}l_1^2$.
However, by rescaling ${\rm d}t=\lambda{\rm d}n$ in eq.~(\ref{de}),
the dependence of $\lambda$ on $|F'(0)|$ can
be removed so that we may assume $\lambda$ {\it is constant}.
This rescaling does not alter the steady state solution of eq.~(\ref{de}).
For growth in radial geometry, the absorber radius of
the walk is $e^{t}\equiv F_1$, while in a cylindrical geometry, the height 
of the walk is $t$.
Once $\lambda$ is held fixed, eq.~(\ref{de}) is the continuum limit
of a growth process in which at each growth step
$F(z)\rightarrow F(f_{0,s_1}(z\pm s_2))$, so that the fixed cutoff in
physical space is replaced by a fixed cutoff $s_1$ in the mathematical plane.
This yields a construction of the growth process as a random combination of 
conformal maps without memory (all maps chosen independently from
one of two possibilities).  This dynamics has a microscopic
conformal invariance, demonstrating the
conformal invariance of the LRW, which could have been anticipated by the
known conformal invariance of the $\eta=1$ LRW on-lattice.  
An image of
a walk in cylindrical geometry produced using this microscopically
conformally invariant
dynamics is shown in Fig.~1, where the density of dots
reveals the variation of the physical space cutoff.

The ansatz $\rho_q=y^{\gamma_1(q)}(1+x^2/y^2)^{\gamma_2(q)}$ solves
eq.~(\ref{de})  with
\be
\label{g1}
\gamma_1(q)=q-2 m \gamma_2(q),
\ee
\be
\label{g2}
\gamma_2(q)=\frac{m+1-\sqrt{(m+1)^2-2 mq}}{2 m},
\ee
where we have chosen the sign of the square-root which gives the dominant
contribution.

{\it Multifractal Spectrum ---}
From the previous section, $\rho_q(0,y)(x,y)$ scales
as $y^{\gamma_1(q)}$ near the growth tip ($x=0$), and
as $y^{\gamma_1(q)-2\gamma_2(q)}$ away from the growth tip.
To derive the multifractal spectrum $\tau(q)$ in physical space
from these, we 
define a multifractal spectrum $\tau_y(q)$ in the mathematical
plane as a preliminary step, 
and then derive a useful inversion relation between $\tau_y(q)$
and $\tau(q)$.

We define $\tau_y(q)$ by
\be
\label{tqd}
\int \frac{{\rm d}x}{y} |y F'(x+iy)|^q
\propto y^{\tau_y(q)}.
\ee
Define a Legendre transform
$f_y(\alpha_y)=q\alpha_y-\tau_y(q)$, with $\alpha_y=\tau_y'(q)$.  
Then dividing the real line into segments of length $y$, the number of
such segments which map under $F$
into points separated by distance $l=y^{\alpha_y}$
in the physical plane scales as $y^{-f_y(\alpha_y)}$.  Conversely,
the number of boxes of size $l$ which map, under the inverse
of $F$, into size $y=l^{\alpha}$, 
scales as $l^{-f(\alpha)}$.  Equating these yields the
inversion relation for $f(\alpha)$
\be
\label{lt1}
\alpha=1/\alpha_y,\qquad
f(\alpha)=\alpha f_y(\alpha_y).
\ee
From eq.~(\ref{lt1}) the functional inversion relation for $\tau(q)$ follows,
\be
\label{funceq}
q=-\tau_y(-\tau(q)).
\ee

We can break the integral in eq.~(\ref{tqd}) into a contribution away
from the growth tip
for $x>>y$, and a contribution due to the growth tip for 
$x\lesssim y$.  The first contribution alone
would give $\tau_y(q)=\gamma_1(q)-2\gamma_2(q) + q-1$, 
while the second
alone would give $\tau_y(q)=\gamma_1(q)+q$.  While the true 
$\tau_y(q)$ is 
obtained by the minimum of these two values, we choose to consider these
contributions separately, giving moments away from the tip and near the tip.

The Legendre transform of $\tau_y(q)$ away from the tip is
$f_y(\alpha_y)=1-(1+m)^2[(2-\alpha_y)+
1/(2-\alpha_y)-2]/(2 m)$.  This is maximum at $\alpha_y=1$ (Makarov's
theorem\cite{makarov}), with $f_y(1)=1$ (Gauss's law).  
The inversion formula (\ref{lt1}) gives
\be
\label{away}
f_{\rm away}(\alpha)=\alpha+\frac{(1+m)^2}{4 m}\Bigl( 1-\frac{1}{2} (2 \alpha-1+
\frac{1}{2\alpha-1})\Bigr).
\ee
This coincides with the CFT spectrum for
$c=13-6m-6/m$.  Thus, $c=1$ implies $m=1$.  For 
self-avoiding walks ($c=0$) $m=2/3$ and for $\eta=1$ on-lattice
LRWs ($c=-2$) $m=1/2$.  For $m<1$, the fractal dimension $D=1+m/2$, so the
maximum dimension is $3/2$.

The Legendre transform of $\tau_y(q)$ near the tip is
$f_y(\alpha_y)=1+m-(2-\alpha_y)(1+m)^2/(2m)-(m/2)[1/(2-\alpha_y)]$.
In this case, the inversion formula (\ref{lt1}) gives
\be
\label{tip}
f_{\rm tip}(\alpha)=
-\frac{m}{8(2\alpha-1)}+\frac{2+m}{4}-\frac{(2+m)^2(2\alpha-1)}{8 m}.
\ee
This has a maximum of zero at $\alpha=(1+m)/(2+m)$.
We find the superuniversal result that for $\alpha<2/3$, $f(\alpha)$ is
dominated by the tip contribution (\ref{tip}), while for 
$\alpha>2/3$, $f(\alpha)$
is dominated by contributions away from the tip.

Finally, for fixed $y<<x$, $\rho_q(x,y)\propto x^{2 \gamma_2}$.
This gives a further set of exponents for decay of the field
along the cluster as a function of distance $x$ from the tip at small $y$, 
which differ from the exponents for
decay of the field at $x=0$ as a function of $y$.
In this case a Legendre transform of $2 \gamma_2(q)$ yields a probability
\be
x^{-\alpha(m+1)^2/2m-1/(2 m \alpha)+(m+1)/m}
\ee
of having a field of order
$x^{-\alpha}$, with $\alpha>0$.  Note that while the most likely value
is $\alpha=1/(m+1)$, it is possible to have fields which are of the same order 
as the field at the tip.  However, there is no scaling relation
to connect these exponents to the decay of the field as a function of
distance from the tip in physical space at fixed small distance away from
the cluster.  The reason we had the inversion relations (\ref{lt1})
above is that at a distance $y$ from the cluster, the field is smooth
on a scale $y$; however, for $y<<x$, the field is not smooth on the scale
of $x$.

The exponents (\ref{away}) are invariant under
$m\rightarrow 1/m$, although the other sets of exponents are not.
For $m\geq 1$ the diffusion term is sufficiently strong
that the growth process does not give rise to linear walks, but
rather clusters, as a point
with $y\approx 0$ and large $x$ can diffuse back to $x=0$ despite the
advection term in eq.~(\ref{de}) (see Figs.~2,3).
While the dimension of such clusters increases above $3/2$ as $m$ increases
above 1, the dimension of the
perimeter decreases as $m$ increases
and remains below $3/2$, giving the invariance of exponents away from
the tip mentioned above.
Unsurprisingly, then,
that the exponents near the tip differ and depend on $m$ and not just on $c$.

{\it Dynamical Scaling---}
There is a form of the electrostatic scaling law\cite{electro} for this system.
We have (in radial geometry) $({\rm d}{n}/{\rm d}F_1)^q=1/(\lambda F_1)^q$.
Taking averages on both sides, and noting that $\langle \lambda^{-q}
\rangle\propto (F_1/l_1)^{2q} (l_1/F_1)^{\tau(-2q)+2q}\propto 
F_1^{-\tau(-2q)}$, 
we get
\be
\label{adt}
\langle({\rm d}{n}/{\rm d}F_1)^q\rangle\propto F_1^{-q-\tau(-2q)},
\ee
where $\tau(q)$ is the scaling near the tip, obtained from Legendre transform 
of eq.~(\ref{tip}).  Now consider
the average mass, $n$, at given radius, $F_1$.  Integrating eq.~(\ref{adt}),
we obtain $\langle n \rangle\propto F_1^{-\tau(-2)}$.  Since
$-\tau(-2)=1+m/2$, for $m\leq 2$, the dimension obtained from the electrostatic
scaling law agrees with that obtained above.  This also gives
a dimension greater than $3/2$ for clusters with $m>1$.
However, eq.~(\ref{adt}) shows that there are large fluctuations in the
growth rate, and that the typical growth is slower than average.

Now we consider the time-averaged growth rate.  After
transforming to time $t$, eq.~(\ref{de}) has only simple scaling in the
mathematical plane,
$t\propto\sqrt{y}$.  
As a first step to finding moments of the time-averaged growth rate,
consider the average of a product of $|F'|$ at different times: $\langle
|\prod\limits_i y_i F'(x=0,y_i,t_i))|^q \rangle$.
Suppose all $|t_i-t_j|$ are of order $y_0^2>>y_i^2$, so that
the derivatives are correlated only 
down to scale $y_0$, and uncorrelated below.  To determine the effect
of the correlations, consider this average for a {\it single} time
but suppose $F'(0,y_0)$ is given,
which imposes boundary conditions on eq.~(\ref{de}).
Then, $\langle |(y F'(0,y) |^q \rangle \propto
(y/y_0)^{\tau_y(q)} |F'(0,y_0)|^{q}$.
Thus, for $r$ points at separate times the ensemble average is
\begin{eqnarray}
\Bigl\langle |\prod\limits_i (y_i F'(0,y_i,t_i)) |^q 
\Bigr\rangle
\propto
\prod\limits_i (y_i/y_0)^{\tau_y(q)}
\langle |F'(0,y_0)|^{qr}\rangle.
\end{eqnarray}
Transforming to the physical plane we find (given that there
is probability $(y_i/y_0)^{-f_y(\alpha_y)}$ of $y_i$ mapping to 
$l_i\propto (y_i/y_0)^{\alpha_y} |F'(0,y_0)|$ and using arguments as above
eq.~(\ref{lt1})
\be
\label{fusion}
\Bigl\langle 
|\prod\limits_i l_i F'(l_i,t_i)|^{-q}
\Bigr\rangle
\propto
y_0^{rq+\tau_y(-r\tau(q))}
\prod\limits_i l_i^{\tau(q)}.
\ee

We can now 
obtain moments of the change in mass $(n_2-n_1)^q$ between
times $t_1,t_2$ by integrating $\int {\rm d}n/{\rm d}t$ and
applying eq.~(\ref{fusion}).
\be
\label{tcor}
\Bigl\langle\bigl(\int\limits_{t_1}^{t_2} {\rm d}{n}/{\rm d}t\bigr)^q\Bigr
\rangle
\propto F_1^{-q\tau(-2)} (t_2-t_1)^{-q+\frac{\tau_y(-q\tau(-2))}{2}}.
\ee
For $t_2-t_1\gtrsim 1$, this is
$\langle(\int\limits_{t_1}^{t_2} {\rm d}{n}/{\rm d}t)^q\rangle
\propto F_1^{-q\tau(-2)}$, so that the total mass at given radius is 
self-averaging.  Over shorter times, eq.~(\ref{tcor})
indicates multiscaling of the mass-radius relation, or 
dynamical multiscaling connecting the two times $t,n$.  The
origin of these large fluctuations is
that the tip may turn and grow towards the center of the cluster rather
than away, so that $F_1$ may increase very slowly with mass.
The fluctuations in $F_1$ are much less in DLA\cite{conf},
where growth occurs over the whole surface;
while $|F'|^{-2}$ may fluctuate locally, the
average of $|F'|^{-2}$ over the cluster has much lower fluctuations.

{\it Conclusion and Renormalization Group---}
We have demonstrated the conformal invariance of the LRW
and calculated exactly the multifractal exponents.
This provides an opportunity to realize clusters of different
conformal field theories as different LRWs,
giving an interesting connection between the iterated conformal map
technique and CFT.  We obtained scaling of harmonic measure both
along the curve and near the growth tip.  Because we have a solvable dynamic 
model, we are able to obtain exponents characterizing the scaling in time.

After rescaling of time,
the growth process was described by random composition of two different
conformal maps.  Other fractal sets such as the Julia set are
obtained by iterating a single conformal map.

It is interesting to note the lack of universality in these results, so that
by changing the ratio $l_2/l_1$ we are able to obtain continuously varying
exponents.  In the on-lattice LRW defined above, if successive particles
are allowed, instead of attaching to the tip, to attach within some distance
from the tip, we expect the dimension to change as the distance is varied,
even with $\eta$ held fixed.  

Recently, it was argued\cite{rg} that
for $\eta\geq 4$ DBM clusters are asymptotically branchless, and hence
one-dimensional.  One may wonder if instead these branchless clusters
are LRWs, as suggested previously\cite{num}, and have dimension
greater than one.  However, the lack of universality in the
LRW implies that even if such DBM clusters with given $\eta$ were to behave as 
LRWs, the resulting dimension need not be the same as that of
an on-lattice LRW defined with the given $\eta$; this $\eta$
may be renormalized.
More importantly, we claim that an asymptotically branchless DBM {\it cannot}
behave as an LRW; if such a cluster were to behave as an
LRW, then the tip would occasionally double back on
itself, so that the point of strongest electric field would not
be at the tip (as shown, the field at large $x$ in an LRW
can be of order the field at the tip), causing a new side branch
to appear and contradicting the assumption of no branching. 
Thus, the possibility of side branching renormalizes the meandering
of the LRW to zero and what results is a single straight branch.
The asymptotic approach of the dimension to one for DBM clusters with
$\eta\geq 4$ is confirmed numerically\cite{num2}.

Formally, one may extend the RG by
noting that the change in growth trajectory induced by a single
tip-splitting event gives a logarithmic correction to the cluster 
size, an additional contribution beyond the correction proportional to branch 
size considered before\cite{rg}.  This yields, at lowest order in
tip-splitting, a log-squared correction to the cluster size, which is 
accounted for by an additional scaling field, representing the 
effective $m$ of the LRW described by the cluster.  However, 
as argued, the possibility of side branching renormalizes the $m$ to zero.  
Thus, we may ignore these log-squared corrections, as done previously.

{\it Acknowledgements---}
I thank R. M. Bradley and V. Hakim for motivating this study, and
E. Ben-Naim for useful comments.
This work was supported by DOE grant 
W-7405-ENG-36.  

\begin{figure}[!t]
\begin{center}
\leavevmode
\epsfig{figure=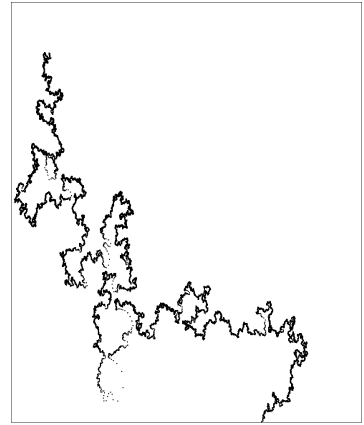,height=8cm,angle=0,scale=.7}
\end{center}
\caption{Plot of an LRW with $m<1$, using conformally invariant microscopic
dynamics and $\sim 50,000$ particles.  
Points indicate position of tip after each growth step.}
\end{figure}
\begin{figure}[!t]
\begin{center}
\leavevmode
\epsfig{figure=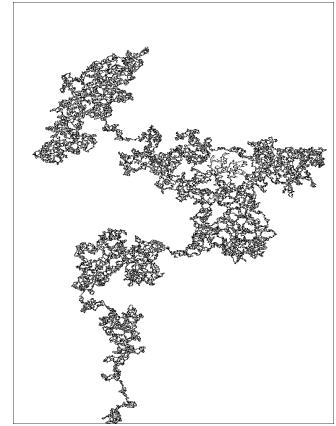,height=8cm,angle=0,scale=.7}
\end{center}
\caption{Plot of an LRW with $m>1$, using fixed cutoff in physical space and
$\sim 50,000$ particles.}
\end{figure}
\begin{figure}[!t]
\begin{center}
\leavevmode
\epsfig{figure=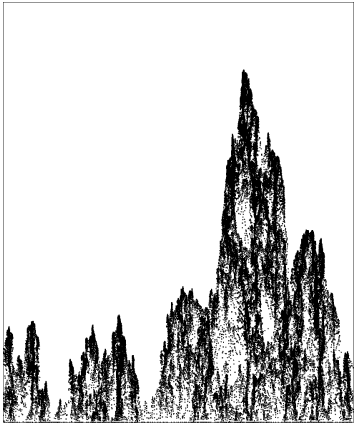,height=8cm,angle=0,scale=.7}
\end{center}
\caption{Plot of an LRW with $m>1$, using conformally invariant microscopic
dynamics and $\sim 50,000$ particles.  $m$ is greater than for the cluster in
Fig.~2.}
\end{figure}  

\vskip-5mm
\begin{thebibliography}{99}
\bibitem{dla} T. A. Witten and L. M. Sander, Phys. Rev. Lett. {\bf 47}, 1400
(1981).

\bibitem{dbm} L. Niemeyer, L. Pietronero, and H. J. Wiesmann, Phys. Rev. Lett.
{\bf 52}, 1033 (1984). 

\bibitem{mf} T. C. Halsey et. al., Phys. Rev. A {\bf 33}, 1141 (1986);
H. G. E. Hentschel and I. Procaccia, Physica D {\bf 8}, 435 (1983); 
T. C. Halsey, K. Honda, and B. Duplantier,
J. Stat. Phys. {\bf 85}, 681 (1996).

\bibitem{makarov} N. G. Makarov, Proc. London Math. Soc. {\bf 51}, 369 (1985).

\bibitem{electro}  T. C. Halsey, Phys. Rev. Lett. {\bf 59}, 2067 (1987).

\bibitem{duplantier} B. Duplantier, Phys. Rev. Lett. {\bf 84}, 1363 (2000).

\bibitem{lrw} J. W. Lyklema, C. Evertsz, and L. Pietronero,
Europhysics Lett. {\bf 2}, 77 (1986).

\bibitem{lrexact} S. N. Majumdar, Phys. Rev. Lett. {\bf 68}, 2329 (1992).

\bibitem{sle} O. Schramm, Israel J. Math. 118 (2000), 221-288; S. Rohde
and O. Schramm, math.PR/0106036.

\bibitem{hl} M. B. Hastings and L. S. Levitov, Physica D {\bf 116}, 244 (1998).

\bibitem{conf} B. Davidovitch et.~al., Phys. Rev. E {\bf 59}, 1368 (1999).

\bibitem{rg} M. B. Hastings, preprint cond-mat/0104344, Phys. Rev. E in press.

\bibitem{num} A. Sanchez et.~al., Phys. Rev. E {\bf 48}, 1296 (1993).

\bibitem{num2} M. B. Hastings, preprint cond-mat/0103312, Phys. Rev. Lett. in
press.
\end{thebibliography}
\end{document}